\documentclass[twocolumn]{aastex631}

\usepackage{siunitx}

\begin{document}

\title{A Contact Binary Mis-Classified as an Ellipsoidal Variable: Complications for Detached Black Hole Searches}

\correspondingauthor{Tyrone~N.~O'Doherty}
\email{tyrone.odoherty@postgrad.curtin.edu.au}

\author[0000-0001-9258-1508]{Tyrone~N.~O'Doherty}
\affiliation{International Centre for Radio Astronomy Research -- Curtin University, GPO Box U1987, Perth, WA 6845, Australia}

\author[0000-0003-2506-6041]{Arash~Bahramian}
\affiliation{International Centre for Radio Astronomy Research -- Curtin University, GPO Box U1987, Perth, WA 6845, Australia}

\author[0000-0003-3441-8299]{Adelle~J.~Goodwin}
\affiliation{International Centre for Radio Astronomy Research -- Curtin University, GPO Box U1987, Perth, WA 6845, Australia}

\author[0000-0003-3124-2814]{James~C.~A.~Miller-Jones}
\affiliation{International Centre for Radio Astronomy Research -- Curtin University, GPO Box U1987, Perth, WA 6845, Australia}

\author[0000-0001-9647-2886]{Jerome~A.~Orosz}
\affiliation{Department of Astronomy, San Diego State University, 5500 Campanile Drive, San Diego, CA 92182, USA}

\author[0000-0002-1468-9668]{Jay~Strader}
\affiliation{Center for Data Intensive and Time Domain Astronomy, Department of Physics and Astronomy, Michigan State University, East Lansing, MI 48824, USA}

\begin{abstract}
Identifying sources exhibiting ellipsoidal variability in large photometric surveys is becoming a promising method to search for candidate detached black holes in binaries. This technique aims to exploit the orbital-phase dependent modulation in optical photometry caused by the black hole distorting the shape of the luminous star to constrain the mass ratio of the binary. Without understanding if, or how much, contamination is present in the candidate black hole samples produced by this new technique it is hard to leverage them for black hole discovery. Here, we follow up one of the best candidates identified from \textit{Gaia} Data Release 3, Gaia DR3 4042390512917208960, with a radial velocity campaign. Combined photometric and radial velocity modelling, along with spectral disentangling, suggests that the true mass ratio (mass of the unseen object divided by the mass of the luminous star) is an order of magnitude smaller than that inferred assuming the modulations arise from ellipsoidal variability. We therefore infer that this system is likely a contact binary, or on the boundary of both stars nearly filling their Roche lobes, however, further observations are required to confidently detect the secondary. We find that the well-known problem of discriminating between ellipsoidal and contact binary light curves results in a larger contamination from contact binaries than previously suggested. Until ellipsoidal variables can be reliably distinguished from contact binaries, samples of black hole candidates selected based on ellipsoidal variability are likely to be highly contaminated by contact binaries or similar systems.

\end{abstract}


\section{Introduction} \label{sec:intro}
The Milky Way Galaxy is expected to contain $10^8-10^9$ stellar-mass black holes (BHs) based on current stellar evolution theories (e.g., \citealt{vandenHeuvel92, Brown94, Timmes96, Samland98, Wiktorowicz19, Olejak20}). However, it is challenging to constrain this population observationally. So far, the only method that has proven successful in identifying isolated BHs is through gravitational microlensing, the first of which was confirmed only recently \citep{Lam22a,Lam22b,Sahu22}. The occurrence of microlensing by isolated BHs is unpredictable, and these events cannot be followed up once the event is over. As a result, studies often focus on looking for BHs in binary systems. Until very recently, the vast majority of confirmed and candidate Galactic BHs came from X-ray binaries (XRBs) (e.g., \citealt{Tetarenko16,Corral-Santana16}). XRBs are binary systems in which a BH accretes from a stellar companion, and in the process radiates liberated gravitational potential energy. These XRBs are primarily identified through X-ray emission arising from accretion. However, broadly, this means the discovery of BHs has been limited to systems in a configuration favourable to accretion. Furthermore, the discovery of transient accreting BHs, which make up the majority, is limited to systems with short recurrence time scales (X-ray all-sky surveys have only been running for several decades). Whilst accreting BHs make up the majority of the known population of Galactic BHs, it is believed that they are only a small component of the total Galactic BH population (e.g., \citealt{vandenHeuvel92,Corral-Santana16}).

Identifying detached BHs in binary systems in the Galaxy has been the focus of numerous studies over the last decade. Of the many initially promising candidates, very few have been found to actually host a BH. While there are $\sim$ 20 accreting Galactic BHs with dynamical mass estimates \citep{Tetarenko16,Corral-Santana16}, there are only 5 confirmed Galactic detached systems with dynamical mass estimates that cannot accommodate a neutron star (NS) in configurations where a non-degenerate companion is infeasible. These are the two BHs discovered in the globular cluster NGC 3201 by a spectral survey \citep{Giesers18,Giesers19}, two BHs in the Galactic field \citep{Shahaf23,El-Badry23a,Tanikawa22,El-Badry23b} initially identified using binary astrometric solutions from \textit{Gaia} Data Release 3 (DR3; \citealt{Gaia,GaiaDR3}), and one system discovered through a spectroscopic survey of Galactic O-type stars \citep{Mahy22}. A sixth possible detached BH is 2MASS J05215658+4359220 \citep{Thompson19,Thompson20}, although while a BH seems likely, the lower limit of the mass estimate ($M = 3.3^{+2.8}_{-0.7}$ M$_\odot$; \citealt{Thompson20}) does not conclusively rule out a NS due to uncertainties in the NS equation of state. A possible seventh detached BH is NGC 1850 BH1 \citep{Saracino22, El-Badry22a, Stevance22, Saracino23}, however, work is still ongoing. This means that in the Galactic field there are only 3, possibly 4, known detached BHs thus far. Developing methods of reliably identifying new BH candidates is critical for discovering more of this elusive population.

The all-sky astrometric, photometric, and spectroscopic \textit{Gaia} mission is providing a wealth of data in which to search for detached black holes. Indeed, two of the confirmed detached BHs in the Galactic field come from an analysis of the astrometric orbits provided with DR3. However, while there are $\sim 2$ billion sources in DR3, there are only $\sim 1$ million binary orbital solutions. There are likely many more sources with detached BH companions that have been observed by \textit{Gaia} that are yet to be identified. While novel techniques have been used to identify promising candidates from \textit{Gaia} (e.g. \citealt{Andrew22,Gomel23}) it is not understood if, or how much, contamination from non-BH systems is present in these samples. Without understanding this possible contamination, it is hard to exploit these new techniques to their full potential. Following up candidates identified using \textit{Gaia} with spectroscopic radial velocity (RV) studies has the potential to substantially increase the known sample of detached BH systems, as well as helping to understand the contamination present in these samples and thereby refine the selection techniques.

Accompanying the release of DR3 were papers detailing the analysis of numerous kinds of variable sources and their classification (see e.g. \citealt[and references therein]{Eyer23,Rimoldini23,Gavras23}). This includes ellipsoidal variables, which were used to identify candidate binaries with compact companions \citep{Gomel23}. In binary systems, the gravity of each component acts on the other, distorting their shape. In BH-star binaries the BH can significantly distort the star into a teardrop shape, resulting in sinusoidal modulations of the observed luminosity of the star on the orbital period.
\citet{Gomel21a} determined a relationship for a `modified minimum mass ratio' (mMMR) that can be computed from the light curve modulation, and does not depend on the mass and radius of the visible star. This mMMR was found to always be less than the minimum mass ratio, which is in turn less than the true mass ratio ($q=M_{\mathrm{BH}}/M_{\mathrm{star}}$). The main physical assumptions underlying the derived relationship are that the primary star fills its Roche Lobe and that the binary has an inclination of 90\si{\degree} (deviations from these assumptions imply the true mass ratio is larger than the mMMR). The other main assumption is that the light is coming from the primary star only. This method was applied to the variable sources identified as ellipsoidal variables by the \textit{Gaia} Variability Pipeline \citep{Eyer23} classifier \citep{Rimoldini23}, producing 6306 detached or weakly-accreting BH and NS candidates.

The 6306 candidates identified by \citet{Gomel23} are a large sample, and if all are real, would represent a substantial increase to the known populations of BHs and NSs. However, as there is likely some level of contamination from systems like contact binaries \citep{Gomel23}, follow up is required. This work builds on that carried out by \citet{Nagarajan23}, who followed up 14 systems from \citet{Gomel23} and found they were unlikely to contain BHs, but did not conclusively characterise the systems. In this paper, we investigate one of the most promising BH candidates from the \citet{Gomel23} sample and discuss the results of our spectroscopic and photometric modelling that reveal it is likely a contact binary. In Section \ref{sec:source_selection} we discuss the selection of this source for targeted follow-up and the details of the spectroscopic campaign. Details of the RV extraction and modelling, joint photometric and RV modelling, spectral disentangling, and the associated results of the analyses are in Section \ref{sec:analysis_and_results}. The results are discussed in Section \ref{sec:discussion}.

\section{Observations}
\subsection{Source Selection}
\label{sec:source_selection}
We began by filtering the 6306 candidates from \citet{Gomel23}, retaining sources for which we could estimate a reliable distance and with no evidence of a luminous companion in \textit{Gaia}. These restrictions manifested as a cut on parallax significance greater than 3 and requiring the \textit{Gaia} parameter \textsc{ipd\_multi\_frac} = 0 (the fraction of successful windows in which their fitting algorithm identified a double peak, and thus potentially a resolved double star, be it a true binary or a visual double). From here, we began exploration by restricting mMMR to be at least $2$, leaving a sample of 13 sources. From this sample of thirteen, there is one particularly interesting source, Gaia DR3 4042390512917208960. 

Gaia DR3 4042390512917208960 has a mMMR of $2.54$ ($1.78$ at one sigma). The mass estimated by the Gaia team for the luminous star by leveraging astrometry, photometry, and spectroscopy is $1.809^{+0.058}_{-0.054}$ M$_\odot$ \citep{Creevey23,Fouesneau23}. Assuming this mass is correct, the inferred mass of the unseen object is $>4.6$ M$_\odot$ ($>3$ M$_\odot$ at one sigma), making it an excellent BH candidate. This source is one of 262 sources with mMMR significantly higher than 1, which were identified by \citet{Gomel23} as being the most promising BH candidates. Furthermore, it is bright ($ G = 13.8$), making spectroscopic follow up feasible with a wide range of instruments.

\begin{figure}[ht!]
\includegraphics[width=\columnwidth]{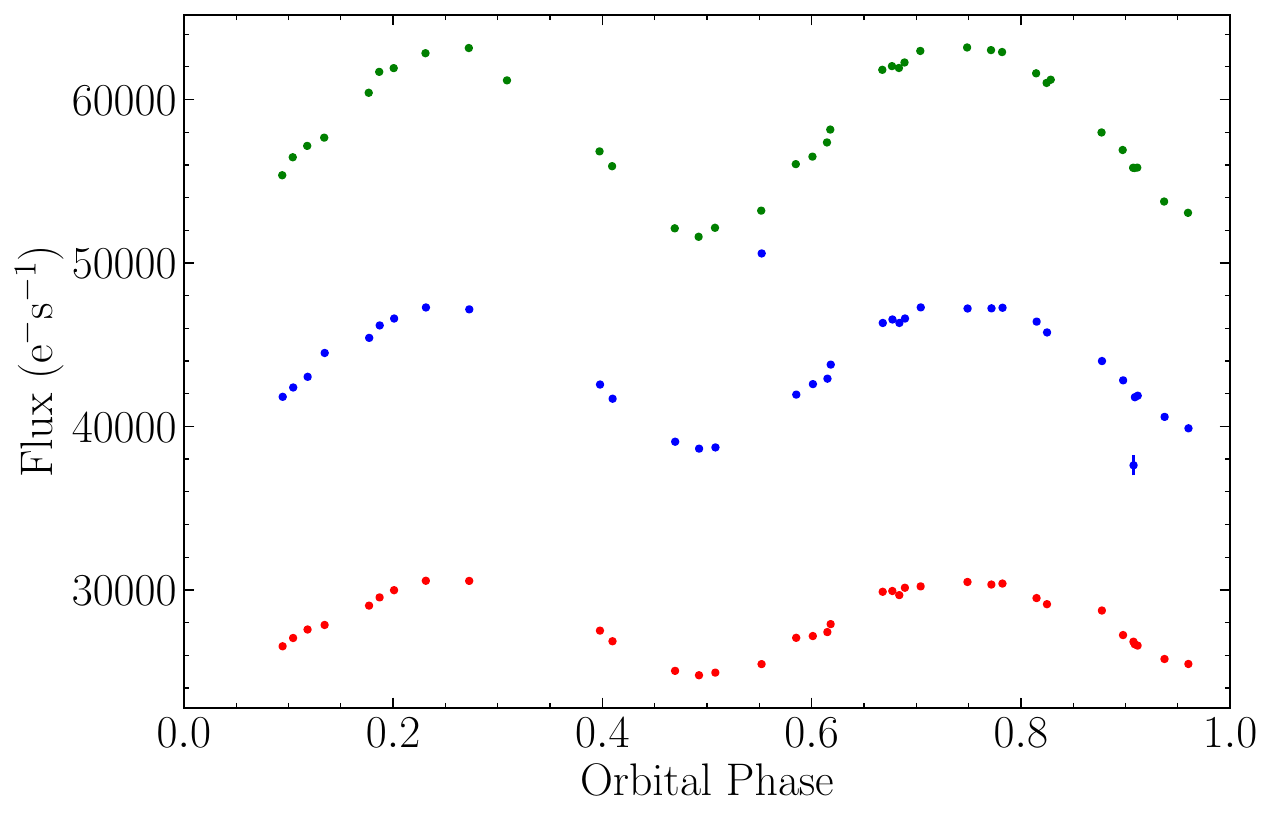}
\caption{Phase folded \textit{Gaia} $G$, $G_\mathrm{RP}$, and $G_\mathrm{BP}$ photometry of Gaia DR3 4042390512917208960. The smooth sinusoidal modulation is evident in all colours.}
\label{fig:gaia_photometry}
\end{figure}

Gaia DR3 4042390512917208960 has excellent astrometry \citep{Lindegren21}, with a parallax significance greater than 20. The renormalised unit weight error (RUWE) for this source is 0.707, suggestive of over-fitting. Other astrometric goodness-of-fit statistics (e.g.\ astrometic excess noise, and all image parameter determination statistics) are as expected for a well-behaved system. A RUWE of 0.707 is therefore likely not a cause for concern. The blue and red photometric excess factor is 1.235, however, this excess is fully consistent with the excess seen in standard sources (figure 18 in \citealt{Riello21}). The \textit{Gaia} $G$, $G_\mathrm{RP}$, and $G_\mathrm{BP}$ phase folded photometry can be found in Figure \ref{fig:gaia_photometry}. The photometry was extracted from the Gaia archive\footnote{\url{https://gea.esac.esa.int/archive/}}. Sinusoidal modulation can clearly be seen in all colours. However, the modulation is not completely smooth, showing some scatter. We think this is likely a result of contamination, which is not unexpected given the source's location in the bulge. Approximately 20\% of the source's transits of the Gaia field of view were marked as blended, ((\textsc{phot\_bp\_n\_blended\_transits} +\textsc{phot\_rp\_n\_blended\_transits})/(\textsc{phot\_rp\_n\_obs} + \textsc{phot\_bp\_n\_obs})), where a blended transit means that there was at least one more source in the observing window \citep{Riello21}. It is important to note that a blended transit occurring is dependent on the scan direction, and thus different transits of the same source can be blended in different ways or not at all \citep{Riello21}. However, the scatter does not significantly alter the general shape of the sinusoidal modulation. In summary, the \textit{Gaia} astrometric and photometric solution is believable and raises no major concerns.

It is worth noting that this source lies slightly above the main sequence (see figure 6 in \citealt{Gomel23}), which could arise from the unseen companion being a luminous object. It is possible to hide a more massive stellar secondary in systems with evolved stars (e.g., \citealt{El-Badry22b}). However, the luminous star in Gaia DR3 4042390512917208960 shows no evidence of being significantly evolved off the main sequence \citep{Fouesneau23}. Furthermore, the estimated mass of the unseen object is large enough such that its luminosity contribution (if it is a main sequence star) would be significant. Cumulatively, as also identified by \citet{Gomel23}, we identified this source as an interesting candidate and worthy of further study.

\subsection{Archival Observations of Gaia DR3 4042390512917208960}
\begin{figure}[ht!]
\includegraphics[width=\columnwidth]{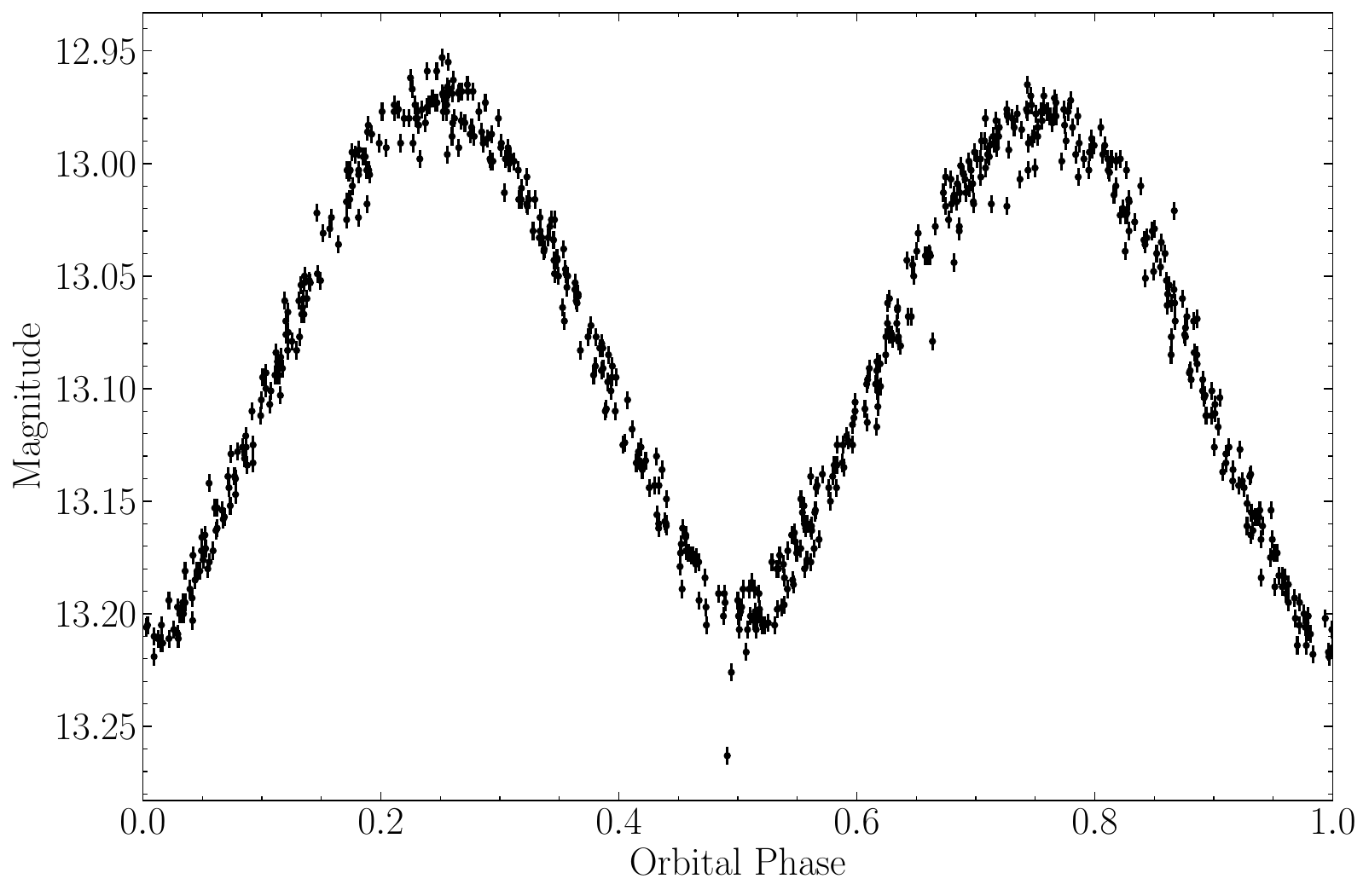}
\caption{Phase folded OGLE I-band photometry. The same shape can be seen as in Figure \ref{fig:gaia_photometry}. }
\label{fig:ogle_photometry}
\end{figure}

The location and brightness of Gaia DR3 4042390512917208960 means it has been observed by both the Optical Gravitational Lensing Experiment (OGLE; \citealt{Udalski15}) and the All-Sky Automated Survey for Supernovae (ASAS-SN; \citealt{Shappee14,Kochanek17}). The OGLE I-band photometry was extracted from the OGLE Collection of Variable Stars online data base\footnote{\url{https://ogledb.astrouw.edu.pl/~ogle/OCVS/ecl_query.php}}, and is presented in Figure \ref{fig:ogle_photometry}. It was classified as an ellipsoidal variable by OGLE (OGLE BLG-ELL-012306; \citealt{Soszynski16}). In their study of OGLE ellipsoidal variables in the Galactic Bulge, \citet{Gomel21b} derived a similar mMMR, also noting it as an interesting candidate. However, the ASAS-SN V team's machine learning classifier classed it as an eclipsing contact binary (ASASSN-V J175613.02-335233.3; \citealt{Jayasinghe20}). Note that ellipsoidal variable was not a classification category in \citet{Jayasinghe20}. Later, \citet{Rowan21} searched for ellipsoidal variables within ASAS-SN data using a combined $\chi^2$ ratio test followed by visual inspection, identifying 369 candidates. However, ASASSN-V J175613.02-335233.3 was not identified as an ellipsoidal candidate in this study. \citet{Gomel23} suggested that it is an ellipsoidal variable based on the \textit{Gaia} and OGLE classification. They also suggested that the comparatively low amplitude of the modulation observed by ASAS-SN arises from contamination as the source is in the Galactic bulge.

\subsection{Dedicated Spectroscopic follow-up}

We obtained spectroscopic observations of Gaia DR3 4042390512917208960 using the Wide Field Spectrograph (WiFeS; \citealt{Dopita07,Dopita10}), an integral field unit (IFU) spectrograph on the Australian National University 2.3 m optical telescope. The spectrograph has both a `blue' and a `red' camera, with which we used the `B7000' and `R7000' gratings ($R \sim 7000$) offering simultaneous wavelength coverage of $4184-5580$ \si{\angstrom} and $5294-7060$ \si{\angstrom}, respectively. Ne-Ar arc lamp exposures were taken immediately following each science observation for wavelength calibration. Data were reduced using the PyWiFeS\footnote{http://pywifes.github.io/pipeline/} pipeline \citep{Childress14}. Data were taken between 2022 June 23 and 2023 April 19 for a total of 24 observations. Individual observations were taken with two 700 \si{\second} exposures (separated temporally only by readout time) so that the duration of each exposure was less than 1\% of the orbit of the binary ($P_\mathrm{orb} = 0.8952134$ d; \citealt{Soszynski16}). Resultant spectra typically have a signal-to-noise of $> 250$ per resolution element in the continuum.

We obtained a single 300 s spectrum of the source with the Goodman Spectrograph \citep{Clemens04} on the SOAR telescope on 2022 October 30. This spectrum used the 2100 l mm$^{-1}$ grating and a 1.2\arcsec\ longslit, giving a full width at half maximum resolution of about 0.9 \AA\ over the wavelength range 6090--6660 \AA. The spectrum was optimally extracted and wavelength calibrated with a time-adjacent arc lamp exposure using standard tools in IRAF \citep{Tody86}, with a signal-to-noise of $\sim 85$ per resolution element in the continuum.

\section{Analysis and Results}
\label{sec:analysis_and_results}

\subsection{Radial Velocity Estimation}
\label{sec:RVE}

The output of the PyWiFeS pipeline is a calibrated data-cube with a spectrum (frequency and brightness) for each pixel in the IFU. We extracted spectra of Gaia DR3 4042390512917208960 from pixels of the WiFeS IFU centred on the source whose total flux was greater than 5 times the root mean square (RMS) flux of background pixels, followed by a subtraction of an averaged background spectrum. To improve signal, for each observation, the two exposures were averaged before proceeding to radial velocity (RV) estimation, leaving us with 24 RVs spread over 10 months.

\begin{figure*}[ht!]
\includegraphics[width=\textwidth]{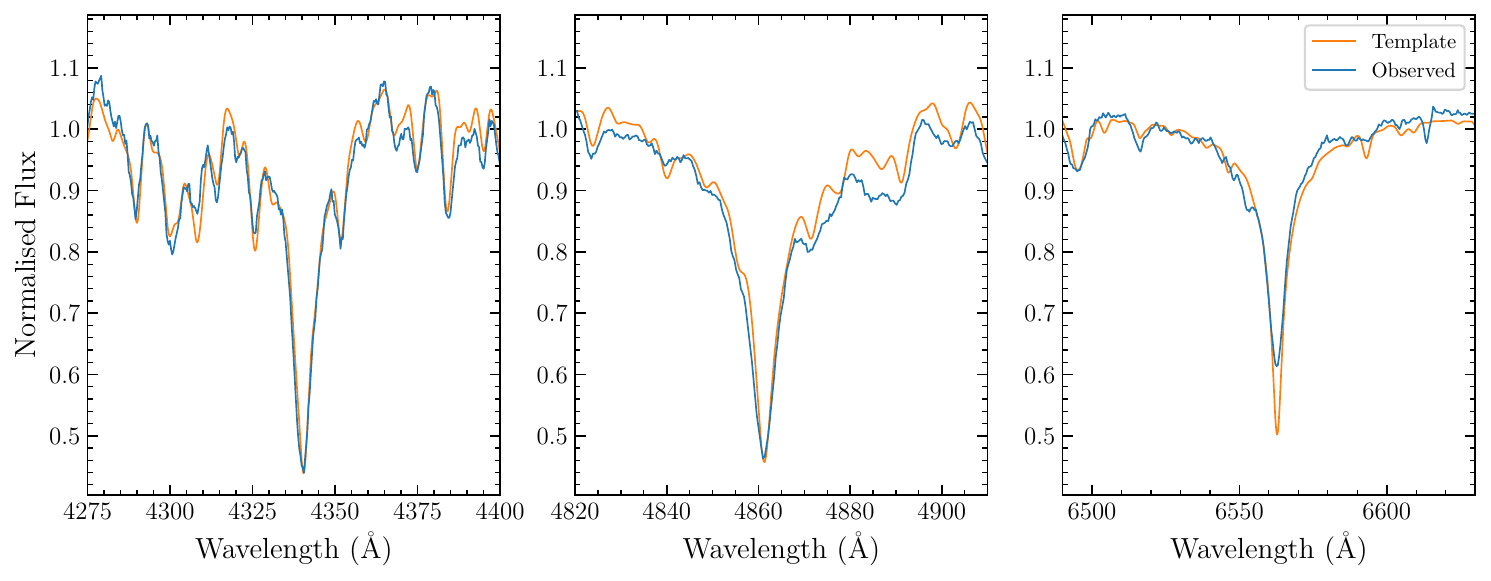}
\caption{Comparison of PHOENIX template ($T_{\mathrm{eff}} = 6700$ \si{\kelvin}, $\log g = 4.0$) spectrum with the shifted and averaged spectrum of all observations of Gaia DR3 4042390512917208960. The left, middle, and right panels are cutouts around H$\gamma$, H$\beta$, and H$\alpha$, respectively. Broadly there is good agreement between the synthetic PHOENIX spectrum and the averaged observed spectrum. Note that the depth of the H$\alpha$ line is much shallower in the observed spectrum.}
\label{fig:template_balmer_series}
\end{figure*}

RVs were estimated via cross-correlation using the IRAF task \textsc{fxcor} \citep{Tody86,Tody93} and a synthetic template spectrum from the PHOENIX library\footnote{http://phoenix.astro.physik.uni-goettingen.de} \citep{Husser13}, which was convolved down to a comparable spectral resolution to the WiFeS data. The library used was the PHOENIX medium resolution grid with uniform grid spacing of 1 \si{\angstrom}. As this spectrum is mainly for determining RVs, and not detailed modelling of the star, we only briefly experimented with metallicity and alpha element abundances, finding no obvious benefit from the use of non-zero values. The effective temperature ($T_{\mathrm{eff}}$) and surface gravity ($\log g$) of Gaia DR3 4042390512917208960 reported in DR3 are $6524^{+31}_{-16}$ K and $3.804^{+0.002}_{-0.003}$, respectively \citep{Creevey23,Fouesneau23}, corresponding to a spectral type of F4V or F4IV. We matched PHOENIX spectra to the WifeS spectra of the target by eye, finding good agreement between the PHOENIX spectrum with $T_{\mathrm{eff}} = 6500$ \si{\kelvin} and $\log g = 4.0$ and the observed spectrum. We found the PHOENIX spectrum with $T_{\mathrm{eff}} = 6700$ \si{\kelvin} and $\log g = 4.0$ to be the best match when considering the depths of the absorption lines. The match between the $6700$ \si{\kelvin} template and observation line depth was almost identical for H$\beta$ and H$\gamma$. However, the observed H$\alpha$ line depth is significantly shallower. Comparisons of the template and observed spectra are shown in Figure \ref{fig:template_balmer_series}.

\begin{figure*}[ht!]
\includegraphics[width=\textwidth]{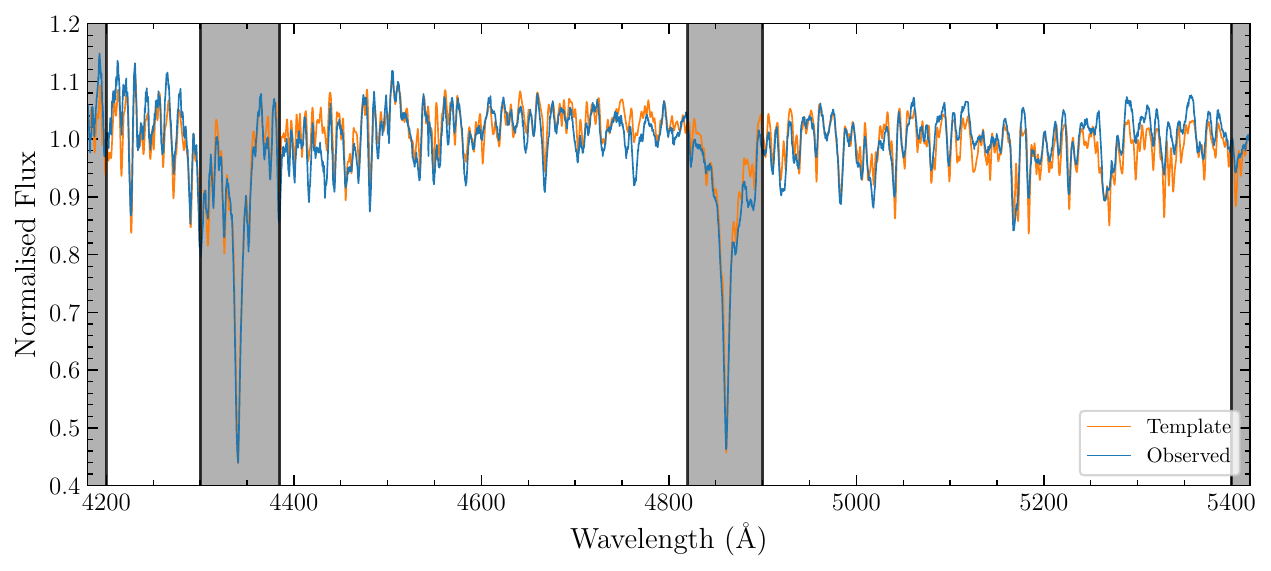}
\caption{Comparison of PHOENIX template ($T_{\mathrm{eff}} = 6700$ \si{\kelvin}, $\log g = 4.0$) spectrum with the shifted and averaged spectrum of all observations of Gaia DR3 4042390512917208960 in the regions used with \textsc{fxcor}. The shaded regions indicate sections of the spectra that were excluded from the cross-correlation. Note that while we exclude H$\gamma$ and H$\beta$, the region includes prominent lines from many atomic species (e.g., Fe, Ca, Na, Mg).}
\label{fig:spectral_region_used_for_cc}
\end{figure*}

We used the `blue' WiFeS spectra for RV estimation due to the abundance of metal lines. Both H$\beta$ and H$\gamma$ were excluded when estimating the RVs to minimise possible contamination arising from emission processes. We used the following regions to estimate RVs with \textsc{fxcor}: 4200-4300 \AA\ , 4385-4820 \AA\ , 4900-5400 \AA\ (Figure \ref{fig:spectral_region_used_for_cc}). All velocities were corrected to the solar system barycenter. A systematic uncertainty of 3 \si{\km \per \second} was added to the uncertainty of each estimated RV \citep{Kuruwita18}. The time of a measured RV was shifted from local observation time to Barycentric Modified Julian Date (BMJD) in the Barycentric Dynamical Time standard (TDB; \citealt{Eastman10}). The RV from the SOAR spectrum was estimated in the same way as the WiFeS RVs using the same PHOENIX spectrum and \textsc{fxcor}. H$\alpha$ was excluded from the region that was cross-correlated to minimise contamination from emission processes. Furthermore, a correction of $1$ \si{\km \per \second} was added based on the telluric oxygen lines in the observation. The estimated RVs for Gaia DR3 4042390512917208960 are presented in Table \ref{tab:RVs}, and the RV from the SOAR spectrum is identified by an asterisk appended to the BMJD. Reported times in Table \ref{tab:RVs} are the mid-observation time.

\begin{table}
    \begin{center}
        \begin{tabular}{cc}
            \hline
            BMJD & RV (\si{\km \per \second}) \\
            \hline
            59781.6833 & $19.3 \pm 9.0$\\
            59810.6610 & $-41 \pm 10$\\
            59753.4166 & $-46.4 \pm 9.2$\\
            59753.4775 & $-29.6 \pm 7.2$\\
            59753.5465 & $-12.5 \pm 6.5$\\
            59753.5947 & $-6.9 \pm 6.5$\\
            59753.6610 & $1.2 \pm 6.6$\\
            59753.6969 & $6.7 \pm 7.1$\\
            59781.4678 & $16.0 \pm 7.4$\\
            59781.5217 & $34.6 \pm 8.2$\\
            59781.5645 & $29.9 \pm 8.2$\\
            59781.6163 & $28.8 \pm 8.6$\\
            59810.4320 & $4.0 \pm 6.4$\\
            59810.4941 & $-7.3 \pm 6.4$\\
            59810.5569 & $-30.0 \pm 7.7$\\
            59822.4157 & $-32.8 \pm 7.5$\\
            59822.4776 & $-11.9 \pm 8.7$\\
            59822.5383 & $-4.3 \pm 6.2$\\
            59840.4369 & $-6.0 \pm 5.9$\\
            59840.4994 & $-4.6 \pm 7.8$\\
            59868.4355 & $34.4 \pm 9.6$\\
            59882.0102* & $-0.2 \pm 9.0$ \\
            60032.7391 & $-43.4 \pm 8.2$\\
            60032.7580 & $-43.7 \pm 7.9$\\
            60053.7026 & $32.8 \pm 9.5$\\

            \hline
        \end{tabular}
        \caption{RVs for Gaia DR3 4042390512917208960 measured using cross-correlation. An asterisk appended to the BMJD indicates the RV was estimated from the SOAR spectrum. The errors in the WiFeS RVs include the 3 \si{\km \per \second} systematic.}
        \label{tab:RVs}
    \end{center}
\end{table}

\subsection{Binary Orbital Parameters}
\label{sec:binary_orbital_parameters}
To fit Keplerian orbits to the estimated RVs we used \textsc{The Joker} \citep{Price-Whelan17}; a custom Monte Carlo (MC) sampler designed for this problem. The parameters we consider are the binary orbital period ($P$), the orbital eccentricity ($e$), the radial velocity semiamplitude of the luminous star ($K$), the systemic radial velocity of the binary ($\gamma$), a (constant) jitter to account for underestimation of RV errors ($s$), the argument of periastron ($\omega$), and the orbital phase at the reference time ($M_0$; by default the reference time used in \textsc{The Joker} is the time of the earliest observation). It is important to note that there may be a systematic offset between the WiFeS and SOAR RVs, but with only 1 SOAR RV we cannot quantify it, and thus we do not attempt to fit for the offset. \textsc{The Joker} samples from the prior probability density functions (PDFs) for each parameter we are considering and then performs rejection sampling, producing samples of the posterior for each parameter.

The priors chosen have significant impact on the rejection sampling when the data is sparse or uninformative and should be considered carefully. The observations of Gaia DR3 4042390512917208960 are quite dense and thus the choice of prior is unlikely to significantly affect the results. Nevertheless, we trialled different priors. We used the default prior implemented in the \textsc{The Joker}, passing values for the the standard deviation of the prior on systemic velocity, $v_0$, and radial velocity semiamplitude, $K_0$, based on the RVs. We trialled different values to asses the impact of these priors on the final results. We used \textsc{The Joker}'s default log-normal prior on period, to which we supplied $P_{min} = 0.25$ d and $P_{max} = 16384$ d as a blind search. We also trialled a much narrower prior centred around the orbital period determined from photometry.

When few samples survive and the posterior appears to be unimodal, the posterior samples from the rejection sampling should be used in standard Markov Chain Monte Carlo (MCMC). We used the No U-Turn Sampling (NUTS; \citealt{hoffman14}) MCMC method to sample the posterior, as implemented in \textsc{pymc3\_ext}. \textsc{pymc3\_ext} contains \textsc{pymc3} \citep{pymc3} extensions extracted from \textsc{exoplanet} \citep{exoplanet}. The priors used here are the same as above, and the results of the rejection sampling are used to initialise the MCMC. Convergence was verified using the Gelman-Rubin diagnostic test \citep{Gelman92}, ensuring the Gelman-Rubin statistic ($\hat{R}$) was close to 1 for all best-fit parameters. We ran 16 chains with 2,000 tune iterations and 8,000 draws, producing 128,000 final draws.

\begin{figure}[ht!]
\includegraphics[width=\columnwidth]{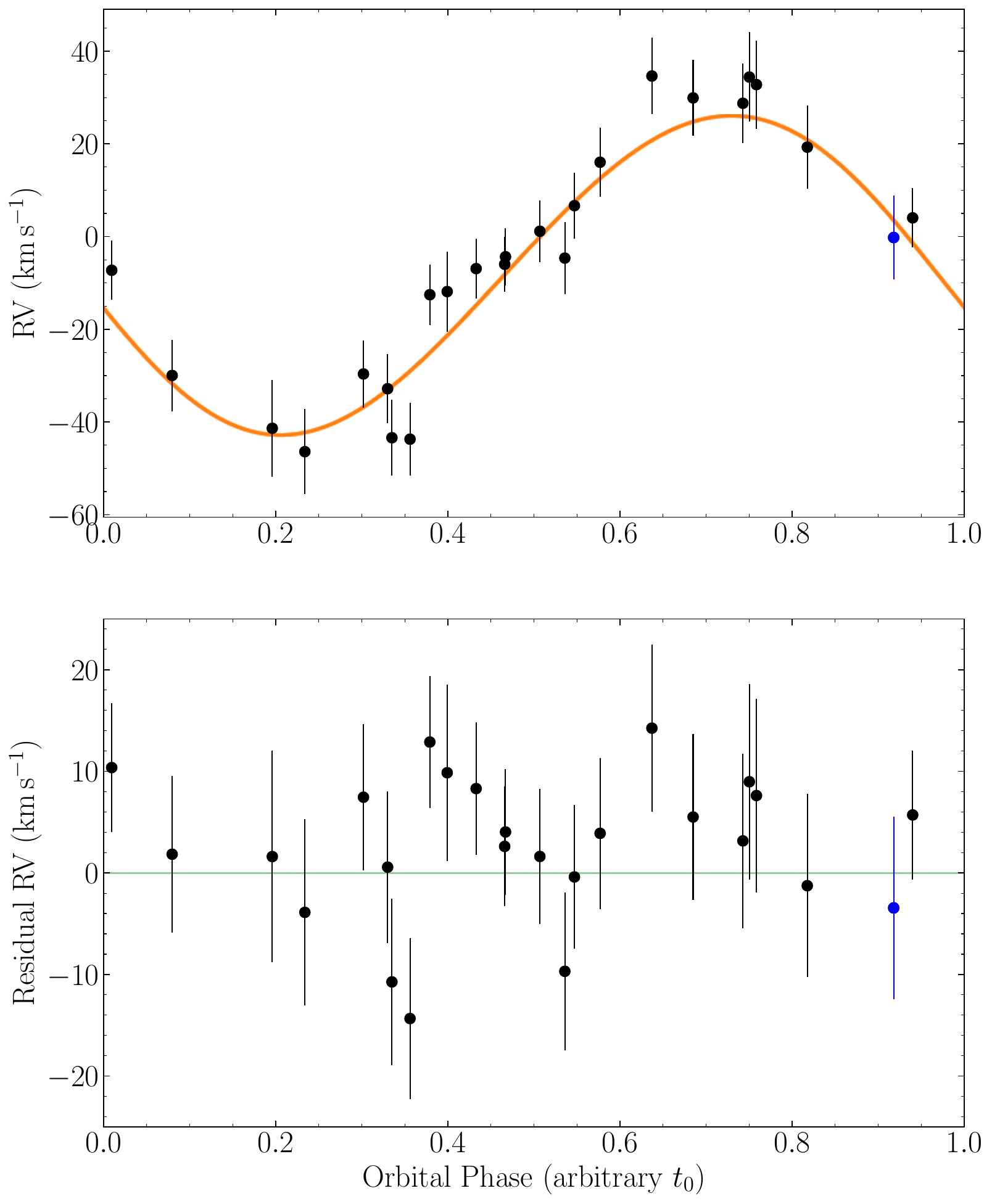}
\caption{Top panel: phase-folded RVs overplotted with the best-fit Keplerian model. Bottom panel: residuals of model minus data as a function of orbital phase. The blue data point is the RV from the SOAR spectrum. The WiFeS RVs are in black.}
\label{fig:orbital_model_residual}
\end{figure}

Changing the standard deviation of the prior on $v_0$ and $K$ had no impact on inferred values, and neither did the choice of period prior. The periods independently inferred from OGLE photometry and WiFeS radial velocities are consistent within error, supporting our belief that the spectra are not affected by significant contamination from nearby stars in the Galactic Bulge. Therefore, we only present one set of results. The binary orbital parameters inferred from \textsc{The Joker} analysis of the RVs are listed in Table \ref{tab:parameters}. While there is some evidence in the posterior for non-zero eccentricity, $e=0$ cannot be excluded. The 95\% upper limit on the eccentricity is 0.18. A comparison between the phase-folded RVs and the best fitting model, with residuals, is shown in Figure \ref{fig:orbital_model_residual}.

There is a noticeable scatter in the RVs visible in Figure \ref{fig:orbital_model_residual}. Additionally, the uncertainty in the RVs determined using \textsc{fxcor} is larger than we would have expected, based on spectra taken with this instrument of other targets. A plausible explanation for both of these is that the radial velocity from a luminous secondary is causing the scatter. However, excluding spectral disentangling (Section \ref{sec:spectral_disentangling}), we found no evidence for the presence of the secondary in the spectra. It is believed that the WiFeS instrument has a degree of inherent instability (see \citealt{Kuruwita18}; standard deviation of the instability is $\sim3$ \si{\km\per\second}), this could also be contributing to the scatter we see. However, the agreement in orbital period separately determined from photometry and \textsc{The Joker} analysis gave us confidence in there being no large systematic issue in the RV modelling.

We then calculated the spectroscopic mass function of the secondary, $\mathbf{m_f}$, using Equation \ref{eq:mf}
\begin{equation}
    \mathbf{m_f =} \frac{M^3_2 \sin^3i}{(M_1 + M_2)^2} = \frac{P K_1^3}{2\pi G} (1 - e^2)^{3/2}
    \label{eq:mf}
\end{equation}
where $G$ is the Newtonian constant of gravitation, $i$ is the inclination, $M_1$ is the mass of the luminous star, $M_2$ is the mass of the unseen object, and the remaining parameters are as described in Table \ref{tab:parameters}. We used the posteriors from the binary parameter estimation to calculate the mass function of the secondary star $\mathbf{m_f =} 3.87^{+0.88}_{-0.77} \times 10^{-3}$ M$_\odot$.

\begin{table}
    \begin{center}
        \begin{tabular}{lc}
            \hline
            Parameter & Value \\
            \hline
            $P$ (d) & $0.89526^{+0.00013}_{-0.00012}$ \\
            $e$ & $0.043^{+0.057}_{-0.033}$ \\
            $K$ (\si{\km \per \second}) & $34.8^{+2.6}_{-2.5}$ \\
            $\gamma$ (\si{\km \per \second}) & $-5.3^{+1.7}_{-1.7}$ \\
            $s$ (\si{\km \per \second}) & $0.0088^{+0.4905}_{-0.0088}$ \\
            $\omega$ (rad) & $2.6^{+2.0}_{-1.4}$ \\
            $M_0$ (rad) & $-0.7^{+2.0}_{-1.4}$ \\
            \hline
        \end{tabular}
        \caption{Orbital parameters determined through modelling the estimated RVs with a Keplerian orbit using \textsc{The Joker}.}
        \label{tab:parameters}
    \end{center}
\end{table}

\subsection{Joint Radial Velocity and Photometry Modelling}
\label{sec:phoebe_modelling}
Now that we have RVs for Gaia DR3 4042390512917208960 we can model them in conjunction with the excellent OGLE photometry. We started jointly modelling the photometry and RVs using the Eclipsing Light Curve (ELC) code described in \citet{Orosz00}. Initial results from ELC suggested the binary was composed of two stars, and that both the stars may be filling their Roche lobes, prompting us to switch to PHOEBE \citep{Prsa05,Conroy20}.
We trialled different binary configurations using PHOEBE, including detached binaries, semi-detached binaries, binaries on the contact binary boundary (both stars are almost filling their Roche lobe), and contact binaries. We used the period and reference time inferred from the OGLE-IV photometry due to the long time baseline of observations. We calculated the flux from magnitudes using the reference wavelength ($\lambda_{\mathrm{eff}}$) for the OGLE-IV I-band filter taken from the Spanish Virtual Observatory (SVO; \citealt{Rodrigo12,Rodrigo20})\footnote{\url{http://svo2.cab.inta-csic.es/svo/theory/fps3/index.php}} assuming OGLE magnitudes are in the AB magnitude system. We selected the logarithmic limb-darkening coefficients of (0.5761,0.2210) from \citet{Claret17} using their radial least squares fit to the log law model for $T_{\mathrm{eff}} = 6500$ \si{\kelvin} and $\log g = 4.0$.

PHOEBE modelling of this system supports the absence of a BH. The large amplitude of the photometric modulation and the low radial velocity semiamplitude of the luminous star are incompatible with the large mass of the unseen object inferred from the photometry using the mMMR. They are also incompatible with a compact object in general. Modelling the system with a neutron star companion (maintaining orbital and stellar parameters constrained from observations) produces a lightcurve with a much lower level of photometric modulation, very different lightcurve minima, and a much larger RV semiamplitude. These cannot be resolved through inclination effects. A larger amplitude of photometric modulation requires a higher inclination, however, decreasing the RV semiamplitude and difference in photometric minima requires a lower inclination. Furthermore, reconciling the observed similarity in photometric minima depths with the high level of photometric modulation is challenging with only a single luminous star due to the physical distortion of the star that is required to recover the large photometric modulation.

When also considering the shape of the lightcurve, both stars must have a high Roche lobe filling factor. This places the system on the boundary between a contact binary and a detached binary with high filling factors. We are not able to discern which is the correct interpretation for this system. The similar depth minima in the lightcurve suggest that each star has a similar temperature. In a contact binary each star typically has a similar effective temperature and in turn likely has a similar spectrum (e.g., \citealt{Pribulla03,Mitnyan20}). Assuming a contact binary configuration, we model the system as a contact binary in PHOEBE to constrain system parameters. An inclination and mass ratio of around $60$\si{\degree} and 0.2, respectively, seem likely. However, in this configuration, there is an apparent contradiction with previous conclusions; the secondary is contributing a significant amount of flux. This would suggest the observed spectra should be double lined and not single lined as they appear.

\subsection{Spectral Disentangling}
\label{sec:spectral_disentangling}
The joint modelling of the photometry and RVs with PHOEBE led us to infer the spectrum of the unseen companion was hidden in the observed spectrum. To test this we explored spectral disentangling, using the shift-and-add approach (e.g., \citealt{Gonzalez06,Shenar20,Shenar22}). When applied to binaries this technique assumes the observed spectra are composites of two individual spectra and exploits that spectral features belonging to each star will move in anti-phase, arising from Doppler shifts. If $K_1$ and/or $K_2$ are unknown and the other orbital parameters are known, a grid-based exploration can be carried out to infer the $K_1$, $K_2$ values that best reproduce the data. This involves evaluating $\chi^2(K_1,K_2)$, the formalism of which is presented in \citet{Shenar22}.

We set the orbital parameters using the period and reference time from OGLE and the systemic velocity from \textsc{The Joker}. Eccentricity was set to 0. This choice is motivated by the indications that the system is a contact binary or on the boundary, and that the RVs determined using the method in Section \ref{sec:RVE} are unreliable due to possible smearing (Sections \ref{sec:phoebe_modelling} and later in this Section). We trialled two different explorations of $K_1$ and $K_2$. The first was holding $K_1$ constant as determined with \textsc{The Joker} and varying $K_2$. The second was varying both $K_1$ and $K_2$ to investigate the possibility of smearing impacting the RVs inferred using the method described in Section \ref{sec:RVE}. The initial guess for this second method was $K_{1,i} = 50$ \si{\km \per \second}, $K_{2,i} = 160$ \si{\km \per \second}, and the grid was split into 100 linearly spaced components between $0.1 K_i$ and $2K_i$. We disentangled the regions around H$\alpha$, H$\beta$, H$\gamma$, $4270-4330$ \si{\angstrom}, and $5150-5200$ \si{\angstrom} as these are the deepest absorption features in the spectrum. We disentangled these regions independently as well as jointly, and also used the entire spectrum as the disentangling region.

\begin{figure}[ht!]
\includegraphics[width=\columnwidth]{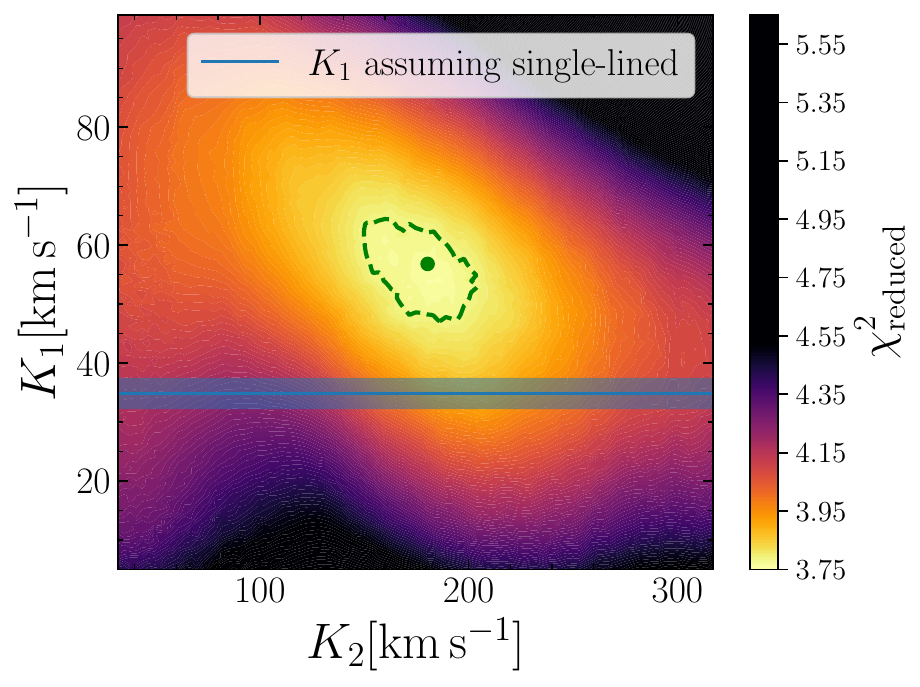}
\caption{$\chi^2$ surface explored in the $K1$, $K2$ grid search, with the green dashed contour indicating 1$\sigma$. The semiamplitudes inferred from this exploration were $K_1 = 54.7 \pm 7.2$ \si{\km \per \second} and $K_2 = 177 \pm 21$ \si{\km \per \second}. The $K_1$ inferred assuming the binary was single lined ($34.8^{+2.6}_{-2.5}$ \si{\km \per \second}) is discrepant with the value inferred using this technique at the $2\sigma$ level.}
\label{fig:chi_squared}
\end{figure}

\begin{figure*}[ht!]
\includegraphics[width=\textwidth]{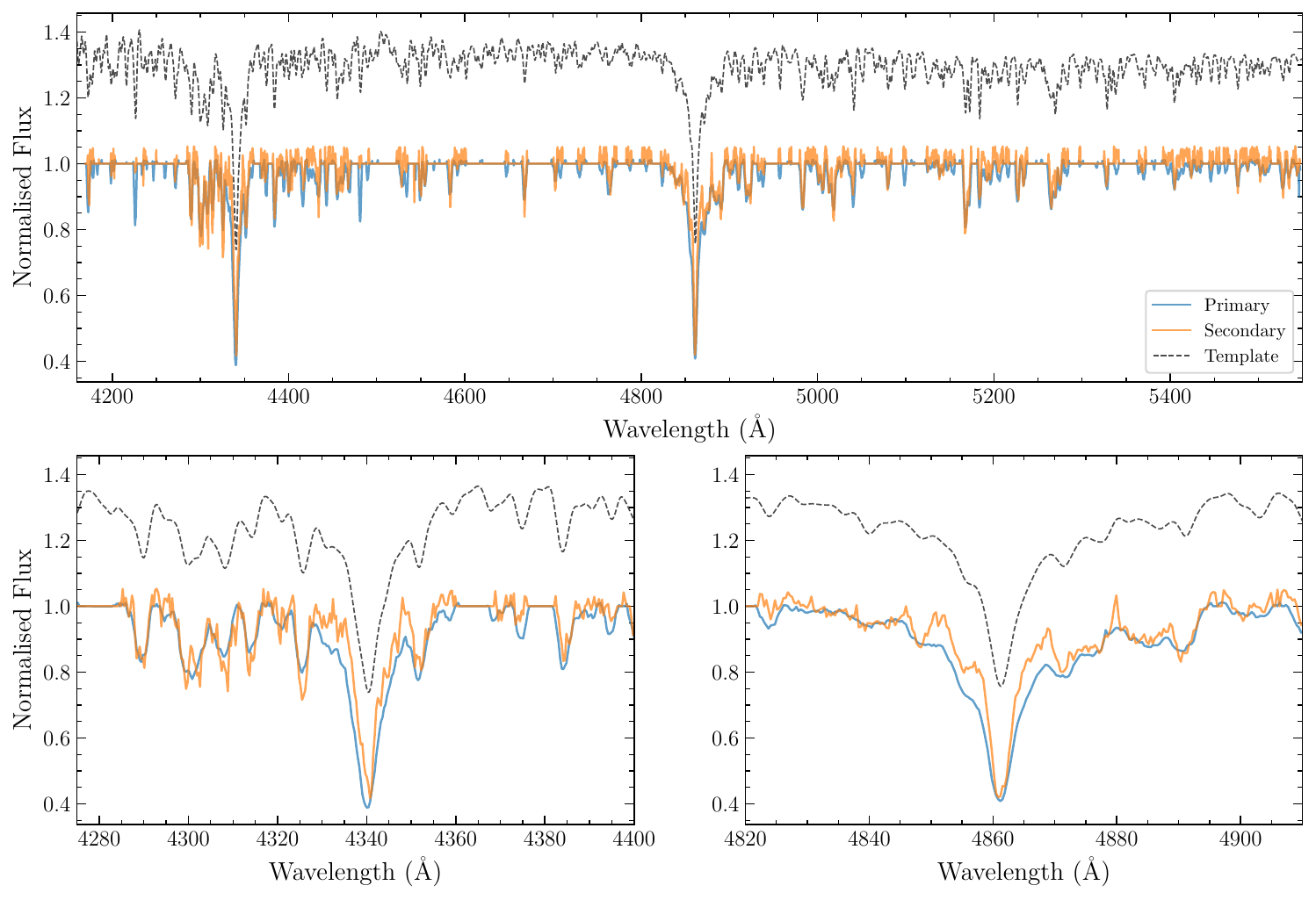}
\caption{Comparison between the disentangled spectra of the components and the PHOENIX template model identified as the best match to the observed data. The secondary has been scaled for a light ratio such that $l_2/l_{\mathrm{tot}} = 0.2$. Noticeably the disentangled spectra are very similar. The lower left and right panels are cutouts around H$\gamma$ and H$\beta$, respectively. In all panels the normalised template has been shifted vertically by 0.3 for clarity. }
\label{fig:disentangled_spectra}
\end{figure*}

Spectral disentangling reveals this system is likely a double-lined spectroscopic binary (SB2), and not a single-lined spectroscopic binary (SB1) as initially thought. Spectral disentangling while holding $K_1$ constant suggested the presence of a luminous companion. However, the features in the disentangled spectrum did not correspond to a physically believable stellar spectrum. Therefore, we only present the results of the grid disentangling where both $K_1$ and $K_2$ were varied. Disentangling the regions as described above return consistent values of $K_1$ and $K_2$ in all tests, and the resulting disentangled spectrum is consistent with a stellar spectrum. Disentangling all regions excluding H$\alpha$ (it is not in the `blue' WiFeS spectra) produced $K_1 = 54.7 \pm 7.2$ \si{\km \per \second} and $K_2 = 177 \pm 21$ \si{\km \per \second}. The $\chi^2$ map is shown in Figure \ref{fig:chi_squared}. The value for $K_1$ inferred from spectral disentangling is discrepant at the $2\sigma$ level with the $K_1$ inferred assuming the binary is single-lined. The disentangled spectrum for each component is presented in Figure \ref{fig:disentangled_spectra} along with the PHOENIX template spectrum. They both appear to have a spectral type of either F4V or F4IV, sharing the same prominent lines and species as the spectra discussed in Section \ref{sec:RVE}.

\section{Discussion}
\label{sec:discussion}
The results inferred from our dedicated spectroscopic campaign and existing photometry imply there is no BH in this system. Considering the RVs in isolation, the system would need to be almost face-on to accommodate a BH. Jointly modelling both RVs and photometry demonstrate that there is no BH in the system, and that the system is more in line with a contact binary or a stellar binary where both stars are almost filling their Roche lobes.

\subsection{Radial Velocity Analysis}
\label{sec:RVA}
Based on the spectroscopic mass function alone, the presence of a BH seems unlikely. If we assume the mass of the luminous star in the binary to be $M = 1.7 \pm 0.05$  M$_\odot$ -- the midpoint between the mass estimated by \textit{Gaia} single star evolutionary models \citep{Creevey23,Fouesneau23} and the mass from the PHOENIX template spectrum -- the minimum mass of the secondary is $0.25^{+0.02}_{-0.02}$ M$_\odot$. This implies that for the secondary to have a mass greater than 3 M$_\odot$ the inclination must be less than $5$\si{\degree}. However, the large photometric variation rules out low inclination angles (Section \ref{sec:phoebe_modelling}). For a conservative lower limit on inclination from the photometry ($i > 50$\si{\degree}), the mass of the primary would have to be greater than 53 M$_\odot$ for the mass of the unseen secondary to be $>3$ M$_\odot$. The presence of a BH is therefore ruled out.

However, it is important to consider the results of spectral disentangling discussed in Section \ref{sec:discussion_spectral_disentangling}. The RVs inferred as described in Section \ref{sec:RVE} are likely smeared due to the presence of a luminous companion in the spectra. This implies that $K$ and $e$ as inferred through this analysis are likely not reliable. In particular, the eccentricity inferred here should not be used as the basis for an argument against the contact binary scenario. If the eccentricity is truly non-zero this argues strongly for a detached system with both stars nearly, but not quite, filling their Roche lobes. 

\subsection{The Single- or Double-lined Nature of the Spectra}
\label{sec:discussion_spectral_disentangling}
Based on the configuration determined using PHOEBE, the secondary is contributing a significant fraction of the light of the system ($>10$\%). However, when visually inspecting the observed spectra we see no evidence for the lines of the secondary. These two conclusions seem mutually exclusive based on the quality of our spectra. 

The results of spectral disentangling suggest the system is an SB2 binary, with the spectrum of the secondary buried in the observed spectra. However, the spectra  of each component are very similar (Figure \ref{fig:disentangled_spectra}). Small errors in $K_1$ can lead to the secondary's disentangled spectrum spuriously mimicking that of the primary \citep{Shenar22}. However, in our case the $K_1$ value determined from spectral disentangling is $\sim60$\% larger than the $K_1$ value determined assuming the spectra are single lined. Furthermore, the disentangled spectrum and the $K_1$ and $K_2$ estimates are consistent regardless of the region of the spectra that is used (as stated in Section \ref{sec:spectral_disentangling}). Assuming the secondary contributes $15-30$\% of the light (motivated by PHOEBE modelling), we simulated synthetic SB2 spectra with similar resolution, sampling, and signal to noise as the observed spectra using the PHOENIX template as the input spectra for both components. Exploring light ratios from 0.0001 to 0.5 supports our results (Appendix \ref{appendix:spectral_disentangling}). Additionally, the magnitude of $K_1$ identified from \textsc{The Joker} is less than the $K_1$ identified from spectral disentangling. This is typical behaviour when the spectra of both components are blended together \citep{Bodensteiner21,Banyard22,Shenar22}. As a result, we are inclined to trust the results of the spectral disentangling analysis. However, we stress that further observations with a higher resolution spectrograph and more stable instrument are required to confirm the detection of the secondary.

That the disentangled spectra are similar is not surprising if the binary is a contact binary, and in turn provides evidence that the system is a contact binary. While this is a circular argument, it is self-consistent, and in line with all other avenues of investigation for this system. For a detached binary with such a mass ratio it would seem unlikely that both components have such a similar spectrum.

With fitted values for $K_1$ and $K_2$ we can determine the mass ratio of the binary. Assuming a circular Keplerian orbit, $q = \frac{K_1}{K_2} = 0.309^{+0.060}_{-0.051}$. Assuming $M_1 = 1.7\pm 0.05$ M$_\odot$ (as discussed in Section \ref{sec:RVA}), then $M_2 = 0.525^{+0.103}_{-0.087}$ M$_\odot$. 

\subsection{Discrepant mass ratios}
\label{sec:discrepant_mass_ratios}

The mass ratio determined from spectral disentangling ($0.309^{+0.060}_{-0.051}$) is significantly different than the mMMR of 2.54 determined by assuming the system is an ellipsoidal variable. The mMMR derived by \citet{Gomel21a} assumes that all the light in the ellipsoidal system is coming from one star, and thus the observed photometric modulation arises due to the distorted star. In the scenario described, and under the assumptions that the star fills its Roche lobe and the binary is observed edge-on, the mMMR is always less than the actual mass ratio. The discrepancy between the mMMR and the mass ratio we determine for Gaia DR3 4042390512917208960 can be explained the same way regardless of whether the system is a contact binary or a binary with both stars being close to filling their Roche lobes. The assumption that all the observed light is coming from a single star breaks down for stellar binaries. In stellar binaries the change in projected surface area as a function of orbital phase is significantly larger than in a star-BH binary due to the presence of two stars. As a result, large modulations can be produced in the light curve purely from the change in projected surface area. As the \citet{Gomel21a} method assumes the light is coming from one tidally distorted star, and the mMMR depends primarily on the amplitude of modulation, binaries with two stars will have inflated mMMRs.

Furthermore, this biases samples selected based on high mMMR towards stellar binaries. A random sample of contact binaries will, on average, have higher mMMRs than a random sample of star-BH binaries with the same broad system parameters. If an initial sample of ellipsoidals are contaminated with stellar binaries, then a sample of high mMMR systems will preferentially include these stellar binaries. The problem is therefore one of classification. While the \citet{Gomel21a} method appears robust, it is only so if it is applied to true ellipsoidal variables. Reducing the contamination should be the priority of future studies applying the \citet{Gomel21a} methodology to identify BHs.

\subsection{Classification of Gaia DR3 4042390512917208960}
\label{sec:classification}
The photometry, spectral disentangling, and PHOEBE modelling we have carried out indicate that Gaia DR3 4042390512917208960 is likely a contact binary, or on the boundary with both stars nearly filling their Roche lobes. Classification of the system is not straightforward from photometry alone. It was incorrectly classified as an ellipsoidal variable by \textit{Gaia} and OGLE, and the initial ASAS-SN variable classification did not have ellipsoidal as a category. However, the system was not identified as an ellipsoidal variable in the later study by \citet{Rowan21} who used a combined $\chi$2 ratio test followed by visual inspection to analyse ASAS-SN data. \textit{Gaia} \citep{Eyer23} and ASAS-SN \citep{Jayasinghe20}
used machine learning classifiers, whereas OGLE \citep{Soszynski16} primarily used template fitting. The templates used by the OGLE team were constructed using the OGLE-IV photometry of bright, previously classified ellipsoidal and eclipsing systems in the best-sampled fields. The solution for better discrimination between contact binaries and ellipsoidal variables from photometry alone is also not straightforward. Whilst mis-classification is often assumed to be due to sparse data, in the case of Gaia DR3 4042390512917208960, simply having more photometric data is not sufficient to resolve the issue. The OGLE-IV I-band light curve for Gaia DR3 4042390512917208960 has 649 data points spanning a time baseline of more than 2000 d (see Figures \ref{fig:gaia_photometry} and \ref{fig:ogle_photometry}), and they also mis-classify the system. However, \textit{Gaia} and OGLE use different methods of classification. Clearly, the classification problem is not limited to one technique, or easily corrected by having more data.

\subsection{Contamination of Ellipsoidal Variable Samples}
The source studied in this work is either a contact binary or on the boundary with both stars nearly filling their Roche lobes. The 14 systems studied by \citet{Nagarajan23} are likely contact binaries, and unlikely to host BHs. It seems likely that the contamination of the \citet{Gomel23} sample with non-ellipsoidal systems is very high. As discussed in Section \ref{sec:discrepant_mass_ratios}, when there is contamination, stellar binaries are likely to dominate in samples selected for high mMMR.

\citet{Gomel23} discussed the difficulty in discriminating between contact binaries and true ellipsoidal systems. They attempted to minimise the contamination by (a) limiting the systems to those with orbital periods $>0.25$ d, and (b) requiring the light curve to have unequal minima. The periods explored by \citet{Nagarajan23} extend to 0.75 d, and the system in this work has a period of 0.8952 d. This suggests that the $P > 0.25$ d restriction is not conservative enough to exclude contact binaries. Furthermore, while unequal minima may help discriminate against other types of variables, it does not exclusively exclude contact binaries and include ellipsoidal variables. Examples of contact binaries with unequal minima include the system studied in this work, systems in \citet{Nagarajan23} if they are truly contact binaries, and also classical contact binary systems such as 44 Bo\"otis, VW Cephei, and Y Sextantis, which can be seen in Transiting Exoplanet Survey Satellite (\textit{TESS}; \citealt{Ricker15}) photometry.

Whilst the mMMR is an excellent technique for identifying candidate compact objects when applied to ellipsoidal variables, the contamination by stellar binaries in the \citet{Gomel23} sample hampers the viability of using the technique in practice. Further studies should prioritise minimising contamination by stellar binaries as much as possible before turning to RV studies. Again, we re-iterate that removing the contaminating binaries is not a simple problem, as discussed in Section \ref{sec:classification}. Furthermore, while for these studies reliable classification is critical, it is also important that the classification is not computationally expensive. For example, \citet{Eyer23} analysed the photometric and spectroscopic time series of 1.8 billion sources from \textit{Gaia} DR3, identifying 10.5 million sources as variable. In the future, the number of known variables will only grow, and so too will the volume of data available (e.g., future \textit{Gaia} data releases, data from new missions such as the Vera C. Rubin Observatory \citep{LSST} and the Nancy Grace Roman Space Telescope \citep{Roman}).

\section{Conclusions}

We have presented a follow-up investigation of one of the most promising BH candidates from the \citet{Gomel23} sample of compact object candidates. This sample was identified by leveraging the relationship between optical light curve modulation amplitude and mass ratio for ellipsoidal variables \citep{Gomel21a} and applying it to sources from \textit{Gaia} DR3. The system we identified as being one of the most interesting candidates from this sample, Gaia DR3 4042390512917208960, had a mMMR of 2.5, implying the unseen secondary was significantly more massive than the luminous star. With detailed multi-epoch spectral observations, we characterised the system, concluding that the binary is unlikely to host a BH. We find the system is most likely a contact binary, with a mass ratio of $0.309^{+0.060}_{-0.051}$, an order of magnitude less than the mMMR. We suggest that this discrepancy arises from the assumption underlying the \citet{Gomel21a} method that all the light comes from one star and is due to ellipsoidal modulation. In contact binaries there are two stars, which gives rise to a significantly larger change in projected surface area over the binary orbit than for ellipsoidal variables with similar binary system parameters. This causes a larger modulation in the optical light curve and results in an artificially large mMMR. This highlights the main issue with leveraging the \citet{Gomel21a} technique to find BHs: without better filtering to remove contact binaries (or other stellar binaries), samples selected based on high mMMR are likely to be heavily contaminated. However, while the contact binary interpretation requires a future study to confirm the detection of the secondary's lines, the rationale for the contamination of the \citet{Gomel21a} sample is still valid for a detached binary in which both stars are close to filling their Roche lobe. Finally, this work highlights the need for dynamical studies to both confirm new BHs and evaluate exciting new techniques for identifying BH candidates.

\begin{acknowledgments}
The authors thank Adela~Kawka, Chris~Lidman, Fiona~H.~Panther, Ian~Price, Kathryn~Ross, Katie~Auchettl, Michael~S.~Bessel, and Michael~Ireland for assistance with WiFeS, Michael~Abdul-Masih, Kyle~E.~Conroy, Sara~Saracino, and Sebastian~Kamann for assitance with PHOEBE, Robin~Humble for assistance with the OzSTAR Supercomputer, and Tsevi~Mazeh and Thomas~J.~Maccarone for helpful discussions. The authors thank the reviewer for their constructive comments that helped improve this work.

T.N.O'D~was supported by a Forrest Research Foundation Scholarship, and an Australian Government Research Training Program (RTP) Stipend and RTP Fee-Offset Scholarship.
This work was supported by the Australian government through the Australian Research Council’s Discovery Projects funding scheme (DP200102471).
We acknowledge extensive use of the SIMBAD database \citep{simbad}, NASA’s Astrophysics Data System, and arXiv. This work has made use of data from the European Space Agency (ESA) mission
{\it Gaia} (\url{https://www.cosmos.esa.int/gaia}), processed by the {\it Gaia}
Data Processing and Analysis Consortium (DPAC,
\url{https://www.cosmos.esa.int/web/gaia/dpac/consortium}). Funding for the DPAC
has been provided by national institutions, in particular the institutions
participating in the {\it Gaia} Multilateral Agreement.
We acknowledge the traditional owners of the land on which the ANU 2.3 m telescope stands, the Gamilaraay people, and pay our respects to elders, past and present.
Based in part on observations obtained at the Southern Astrophysical Research (SOAR) telescope, which is a joint project of the Minist\'{e}rio da Ci\^{e}ncia, Tecnologia e Inova\c{c}\~{o}es (MCTI/LNA) do Brasil, the US National Science Foundation’s NOIRLab, the University of North Carolina at Chapel Hill (UNC), and Michigan State University (MSU).
Part of this work was performed on the OzSTAR national facility at Swinburne University of Technology. The OzSTAR program receives funding in part from the Astronomy National Collaborative Research Infrastructure Strategy (NCRIS) allocation provided by the Australian Government, and from the Victorian Higher Education State Investment Fund (VHESIF) provided by the Victorian Government.
This research has made use of the Spanish Virtual Observatory (https://svo.cab.inta-csic.es) project funded by MCIN/AEI/10.13039/501100011033/ through grant PID2020-112949GB-I00.

\end{acknowledgments}

\vspace{5mm}
\facilities{Gaia, ATT, OGLE, SOAR}

\software{\textsc{aladin} \citep{aladin1,aladin2}, 
\textsc{Astropy} \citep{Astropy},
\textsc{emcee} \citep{emcee},
\textsc{exoplanet} \citep{exoplanet},
\textsc{ELC} \citep{Orosz00},
\textsc{iPython} \citep{ipython}, 
\textsc{IRAF/fxcor} \citep{Tody86,Tody93}, 
\textsc{Matplotlib} \citep{matplotlib}, 
\textsc{Numpy} \citep{numpy}, 
\textsc{Pandas} \citep{pandas}, 
\textsc{PHOEBE} \citep{Prsa05}, 
\textsc{PyMC3} \citep{pymc3}, 
\textsc{SAO DS9} \citep{ds9}, 
\textsc{The Joker} \citep{Price-Whelan17}, 
\textsc{Scipy} \citep{scipy}.}

\appendix
\section{Spectral Disentangling}
\label{appendix:spectral_disentangling}
Disentangling was performed as described in Section \ref{sec:spectral_disentangling}. The SB2 synthetic data were generated using the PHOENIX template spectrum as the model for each star, reflecting the similarity between the disentangled spectra for Gaia DR3 4042390512917208960. 24 synthetic observations were created to reflect the number of observations of the source, and the model spectra were convolved to a resolution and sampling factor similar to the WiFeS data. The signal-to-noise of the synthetic spectra were set so such that the synthetic data were similar to the WiFeS spectra. The same orbital period as Gaia DR3 4042390512917208960 was used, an eccentricity of 0, and $K_1 = 50$ \si{\km \per \second} and $K_2 = 160$ \si{\km \per \second}. The six light contribution fractions ($l_2/l_{\mathrm{tot}}$) examined were 0.5, 0.3, 0.2, 0.1, 0.05, and 0.0001 (reflective of no luminous companion). The $K_1$, $K_2$ grid that was explored was quite coarse; split into 15 linearly spaced components between $0.1 K_i$ and $2K_i$. For this test, $K_{1,i} = 60$ \si{\km \per \second}, $K_{2,i} = 192$ \si{\km \per \second} to shift the centre of the explored grid away from the correct values.

\begin{figure*}[ht!]
\includegraphics[width=0.9\textwidth]{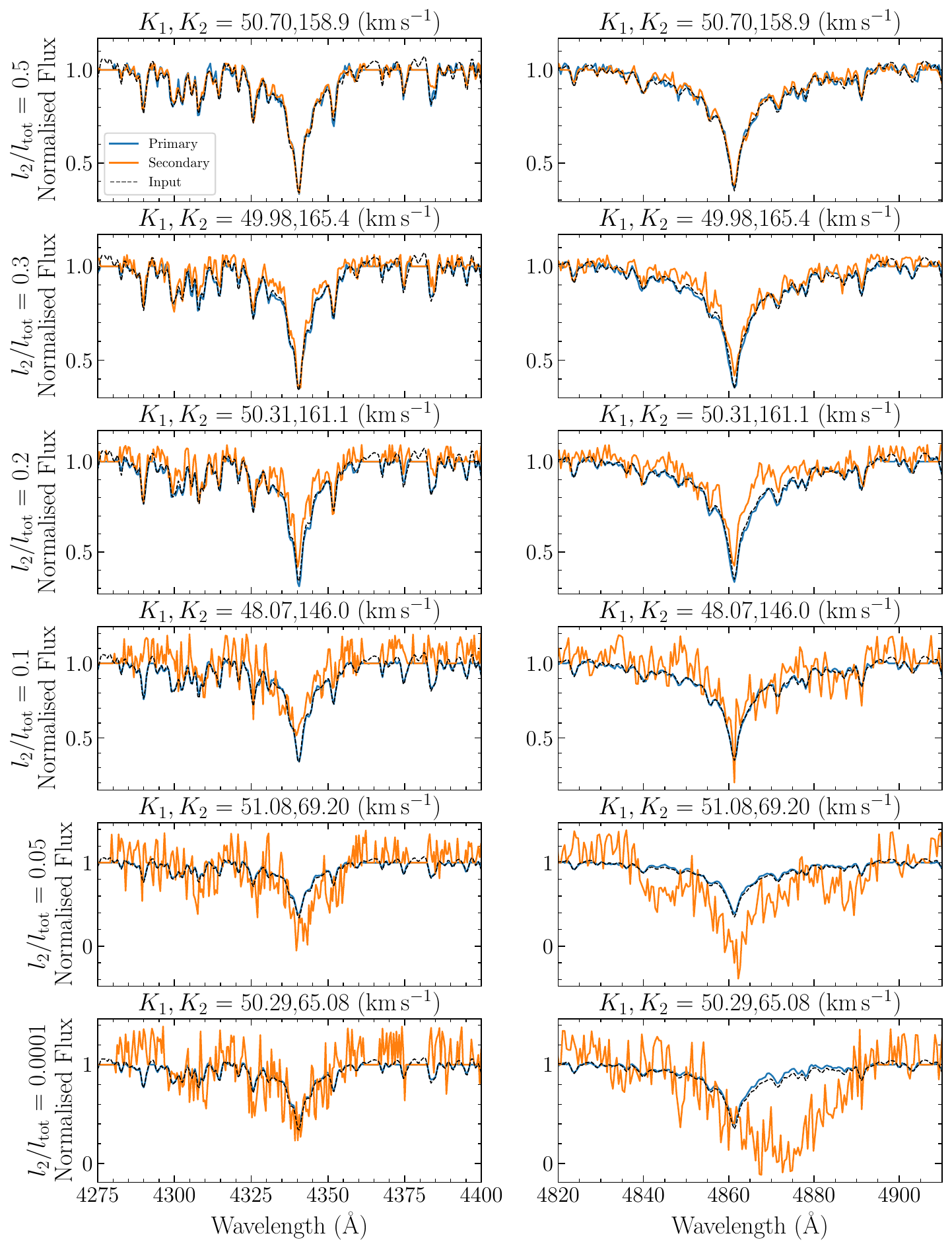}
\caption{Comparison of disentangled spectra with the input spectrum around H$\gamma$ and H$\beta$ for six different light contribution fractions. All secondary spectra are scaled by their light contribution except the bottom row which is scaled assuming $l_2/l_{\mathrm{tot}} = 0.05$. These results suggest that we could reliably detect the presence of a luminous companion if it was contributing at least 20\% of the total light of the system.}
\label{fig:light_ratio_tests}
\end{figure*}

The results presented in Figure \ref{fig:light_ratio_tests} suggest we could reliably constrain $K_2$ and the general shape of the secondary's spectrum when it is contributing $>20$\% of the total light when both components have the same spectrum. At light contribution fractions of 0.1, 0.05, and 0.0001, the disentangled $K_2$ is spurious and the disentangled spectrum of the secondary can mimic the primary. The $\chi^2$ surface for light contribution fractions of 0.1, 0.05, and 0.0001 is not smooth and has no clear minima, whereas for light contribution fractions of 0.2, 0.3, and 0.5 the surface is smooth with a clear minimum. The smooth $\chi^2$ map for Gaia DR3 4042390512917208960 (Figure \ref{fig:chi_squared}) resembles an accurate disentanglement far more than any of the spurious results described in this section. While not conclusive in isolation, we believe it is another piece of supporting evidence that our disentangling results are accurate.

\bibliography{main}{}
\bibliographystyle{aasjournal}

\end{document}